\documentclass[runningheads]{llncs}
\usepackage{graphicx}
\begin{document}
\title{Boost Blockchain Broadcast Propagation with Tree Routing
\thanks{Supported by Brahma OS.}}
%
%
\author{Jia Kan\orcidID{0000-0001-5922-9502} \and
Lingyi Zou\orcidID{0000-0002-6738-4980}\and \\Bella Liu\orcidID{0000-0002-2072-1089}\and Xin Huang\orcidID{0000-0002-1668-9696}}
\institute{XJTLU, Suzhou, China\\
\url{http://www.xjtlu.edu.cn/}}
\maketitle              
\begin{abstract}
In recent years, with the rapid development and popularization of BitCoin, the research of blockchain technology has also shown growth. It has gradually become a new generation of distributed, non-centralized and trust-based technology solution. However, the blockchain operation is expensive and transaction is delayed. Take BitCoin as an example. On the one hand, a block is produced every ten minute. On the other hand, once the new block is generated, it takes a certain time to propagate world wide. The slow speed of propagation determines that BitCoin can not use too small block interval time. Ethereum also faces similar problems, so the concept of uncle block was introduced to reduce blockchain forks. This paper introduces a new tree structure based broadcast propagation routing model, providing a novel method to organize network nodes and message propagation mechanism. In oder to avoid the single node failure problem, the tree cluster routing is proposed. The research shows that the tree based routing can accelerate broadcast convergence time and reduce redundant traffic.

\end{abstract}
\section*{Introduction}

The concept of blockchain has risen rapidly, it has gradually become a hot spot of technological innovation independent of BitCoin\cite{ref_article2}. It creating a new distributed data storage technology with an innovation change on system and programming concepts\cite{ref_book6}. Many financial institutions and related IT enterprises around the world have set off a blockchain technology in the economic and Internet fields. The development of blockchain has gone through three stages\cite{ref_book14}. Blockchain 1.0 era is a cryptocurrency represented by BitCoin, which has the functions of payment, circulation and other currencies. The Blockchain 2.0 era subverts traditional currency and payment concepts through smart contracts, such as Ethereum. In the era of blockchain 3.0, it goes beyond the financial field and provides decentralized solutions for various industries, including education, health, culture, art and so on. Blockchain technology will change the deep structure of the enterprise, create new business model, and ultimately bring huge economic impact.

~\\The blockchain network is a decentralized peer-to-peer (P2P) network that breaks the traditional Client/Server (C/S) model. In the original C/S model network
\cite{ref_book15}, the server requirements are too high. It is increasingly difficult to provide satisfactory service performance. On the contrary, the decentralization of P2P technology is fully compatible with Internet protocol and structure. It has strong adaptability and network service capabilities. In recent years, with the broadband of users and the improvement of computer capacity, the advantages of P2P technology can be fully utilized.

~\\Bitcoin's network uses a P2P network architecture which was based on the Internet, gossip protocol is used to propagate transaction and block information. In 1987, the gossip protocol was first proposed in Epidemic algorithms for replicated database maintenance.\cite{ref_article1}. The gossip protocol is also known as the epidemic algorithm. In a bounded network, each node communicates randomly with other nodes. After some messy communication, the state of all nodes will be agreed\cite{ref_article10}. Figure 1 simulates the case where a message is propagated using the gossip protocol in the entire network nodes: The source node that starts the message will randomly select a peer node to send the message, then the infected node continues to select some peer node propagation. Repeat the process until all nodes are infected, until all nodes receive the message. The shortcomings of gossip are also obvious. In principle, mutual infection between nodes leads to repeated infections of some nodes. Although the reliability of information dissemination is ensured, duplicated contact affects the speed of network propagation convergence, especially for new block discovery. In term of speed, the slowness of propagation affects the efficiency to form a consensus\cite{ref_article7}.

~\\In the blockchain protocol, after block data is generated, the node is broadcasting to all other nodes on the entire network for verification. However, due to the irregular connection between nodes, the message will propagated repeatedly, which ensures the reliability of information, but also causes redundancy and affects the broadcasting efficiency.In a large amount of nodes in the network, it is naturally desired to have a broadcast mechanism that more closely matching this mode to optimize the transmission of blockchain technology related services\cite{ref_article4}. This paper designs a new tree structure model that supports blockchain broadcast services, proposing a complete information broadcast mechanism to reduce redundant traffic within the network and accelerate information propagation.

\section*{Blockchain and Broadcast Network}

Microscopically, the essence of blockchain is a hash chain that cannot be tampered and traceable. Macroscopically, blockchain is a basic protocol with the characteristics of distributed storage, P2P network and consensus mechanism\cite{ref_article2}. The reason why the P2P network is adopted is the blockchain nodes are characterized by openness, autonomy and anonymity.

~\\In the traditional centralize mode, it's common to use a dedicated server from which multiple clients obtain data. The advantage of this mode is that the system is easy to manage and understand. However, the shortcomings of this model are also obvious: (1) due to the centralize mode, the system is prone to a single point of failure. (2) a single server faces a large number of clients, because CPU capacity, memory size and network bandwidth limitations, the number of clients that can be served concurrently is very limited. P2P technology is a network proposed to solve these problems\cite{ref_article9}. In a P2P network, each node can either receive services from other nodes or provide services to other nodes. In this way, huge terminal resources are utilized, two drawbacks in the centralize mode are solved in one fell swoop.

~\\Each node in a P2P network communicates and interacts with each other in a flat topology\cite{ref_article9}. There are no special nodes and hierarchical structures. Each node will assume network routing, verify block data, broadcast block data, discover new nodes, etc. Because there are no special nodes in the whole network, he failure of any node will not pose a threat to the stability of the whole network.Different organizational structures are applied in different fields of computer network.In its network structure organization, such as gossip protocol, plays a huge role in BitCoin.

\section*{Gossip Protocol}

In 1972, the emergence of the simple branch processing model of Galton-Watson made the research on gossip a solid theoretical tool\cite{ref_book13}. The publication of "Epidemic algorithms for replicated database maintenance"\cite{ref_article1} in 1978 pushed the research on gossip to a new height. The gossip protocol is simple and efficient, it also has good scalability and robustness. It is well adapted to a non-central, large-scale, highly dynamic distributed network environment, making it widely used in many fields\cite{ref_article12}.

~\\The gossip protocol is also known as the epidemic algorithm. In a bounded size network, each node randomly communicates with other nodes. After some messy communication, the state of all nodes will be agreed. Each node may know only a few neighbor nodes. As long as these nodes are connected through the network, eventually their state is consistent.

~\\In the gossip protocol\cite{ref_article10}, each node has the information about its neighbors. In every round of communication, each node selects one from the contacts to communicate, three communication methods as following:
1) Push, the A node sends the information to the Node B, Node B updates according to the received information;
2) Pull, Node B sends the information to Node A, Node A updates according to the received information;
3) Push/pull, in addition to the pull, A pushes the data that Node B does not have it, node B updates according to the received information.

~\\There are three important factors in the gossip network\cite{ref_article1}. (1) The number of nodes in the network, that means the size of the network. (2) The number of contacts saved by each node. In layman's terms, each node knows several nodes (number of friends). If the total number of nodes is 10000 and the number of friends is 100. The received message was transmitted multiple times though 100 friends, finally 10000 nodes received this message. (3) Rumor Mongering\cite{ref_article11}, the interest of rumors will be reduced with the same rumors received many times. When hearing this message many times, the node will instinctively believe that the rumor has spread widely, thus stopping its own spread. In 1/k, Increasing k will decrease the residue. This attenuation of interest helps to reduce rumor propagation, thereby reducing bandwidth usage, but also to some extent leads to the formation of isolated points.

~\\Figure 1 simulates the case where a message is propagated using the gossip protocol in the entire network node: The source node that starts the message will randomly select a peer node to send the message, then the infected node continues to select some peer node propagation. Repeat the process until all nodes are infected, that is, all nodes receive the message. As shown in the figure 1,regardless repeating friends, in the first round, node 1-2; in the second round, node 1-3 and node 2-3; third round, node 1-4, node 2—4, node 3—4. It is known that its propagation amount is $2^{N - 1}$. However, in the actual simulation, nodes will have duplicate contacts. As soon as it reaches the boundary, most of the contacts who is spreading with may already know the message. In order to avoid the waste of bandwidth, the interest of mongering will decrease by the times of the contacts heard the news, in the case of the contacts amount of k. At the end of the spread,the feature of reduced interest does not guarantee 100\% coverage.

\begin{figure}
\includegraphics[width=\textwidth]{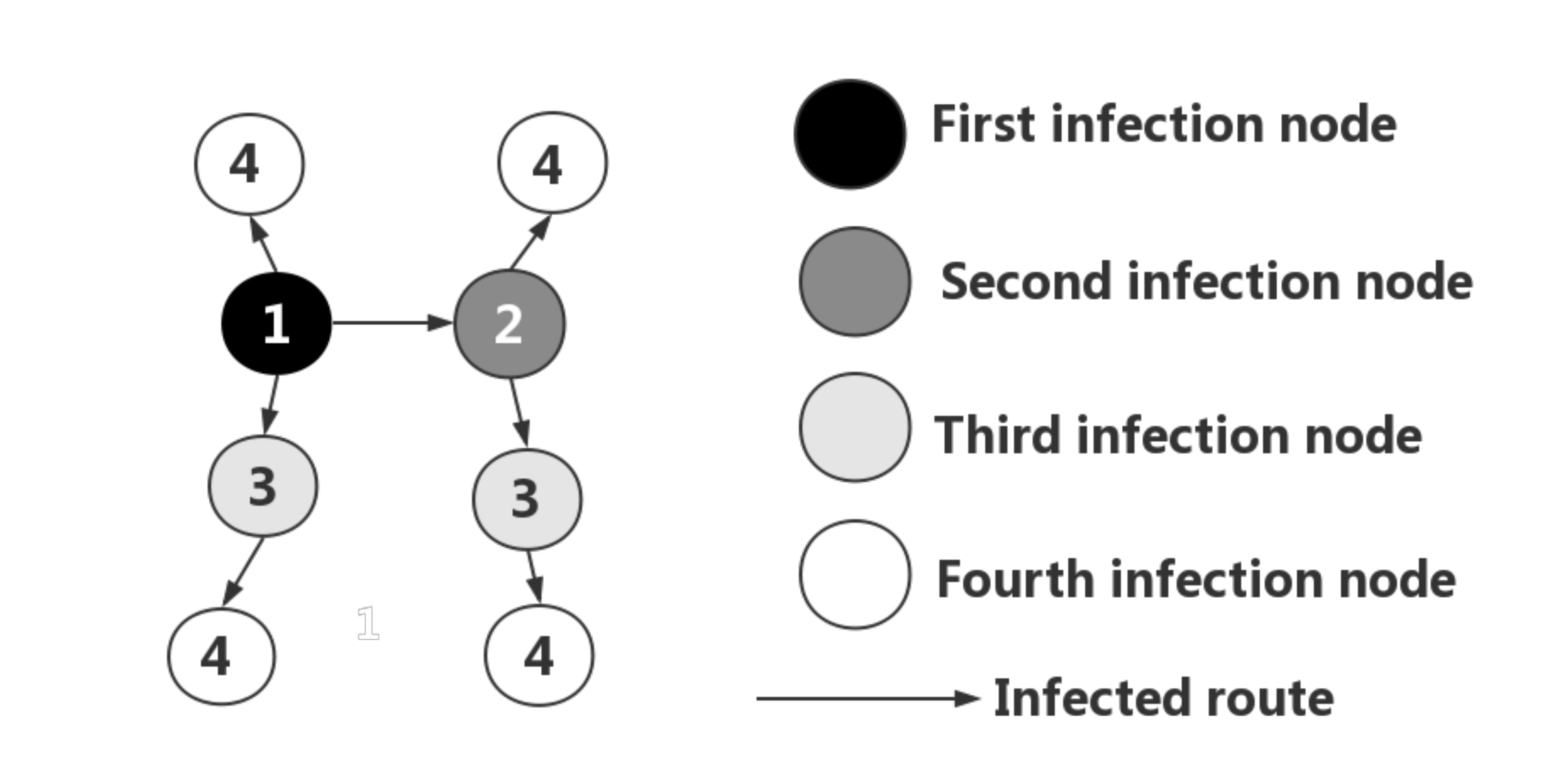}
\caption{The gossip protocol is used to disseminate messages in the network.} \label{fig1}
\end{figure}

~\\The advantages of gossip are also obvious. (1) Any node can spread the message to the whole network. When the central node has an error, the rectification system will not affects. (2) Any node can join or quit. (3) Fault tolerance, the downtime and restart of any node in the network will not affect the spread of messages. (4) Low CPU overhead and low network bandwidth are suit for large P2P networks. Figure 1 simulates the case where a message is propagated using the gossip protocol in the entire network node: The source node that starts the message will randomly select a peer node to send the message, and then the infected node continues to select some peer node propagation. Repeat the process until all nodes are infected, that is, all nodes receive the message. As we can see, we also clearly see the shortcomings of the gossip network. Since the node will only send messages to several nodes randomly, the message will eventually reach the whole network through multiple rounds of dissemination. Some messages have to go very long. In order to reach the far node in the corner of the network, it will inevitably cause delay in the message. In addition, mutual infection between nodes leads to repeated infections of some nodes.It ensures the reliability of information dissemination, but also creates redundancy of information and increases the processing pressure on nodes.

\section*{Tree based broadcast network}

~\\Inside blockchain, P2P network is a mandatory component. Among the three types of P2P communication models: pair-wise, group-wise and broadcast, broadcasting is the most common requirement for blockchain as every transaction or new block discovery requires to be announced in the whole network as efficiency as possible.

~\\The broadcast network can be implemented with gossip protocol\cite{ref_article1}. Gossip protocol\cite{ref_article1} is extreme reliable, however it can not ensure the message can be arrived in every corner of network within fixed time. The propagation convergence time could last very long.

~\\We proposed a tree based network in broadcasting use case. In tree based  network topology, node join to tree as a leaf one by one, rules are set to ensure the tree is as balance as possible. In tree network, the longest distance between two nodes will be less than 2 times of tree height. That means even in a binary tree based topology, the message can be passed to the farthest node with less than 2 times tree height hops. It might be the most efficiency way for broadcasting in logic.

~\\Compared with Figure 1, Figure 2 simulates the case where a message is broadcasting through the tree structure throughout the network node: the source node that starts the message will broadcast the message along the branch to each node, assuming the depth of the source node is n,  This infected node continues to propagate to the next deep (n+1) node until all nodes receive the message. As shown in the figure 2, without considering Rumor Mongering, in the first round, broadcasting depth n, broadcasting path node 1-2; in the second round, broadcasting depth n+1, node 2-3; in the third round, The propagation depth is n+2, 3-4. It can be seen that the rate of its propagation is 1+2+4+8.

\begin{figure}
  \includegraphics[width=\textwidth]{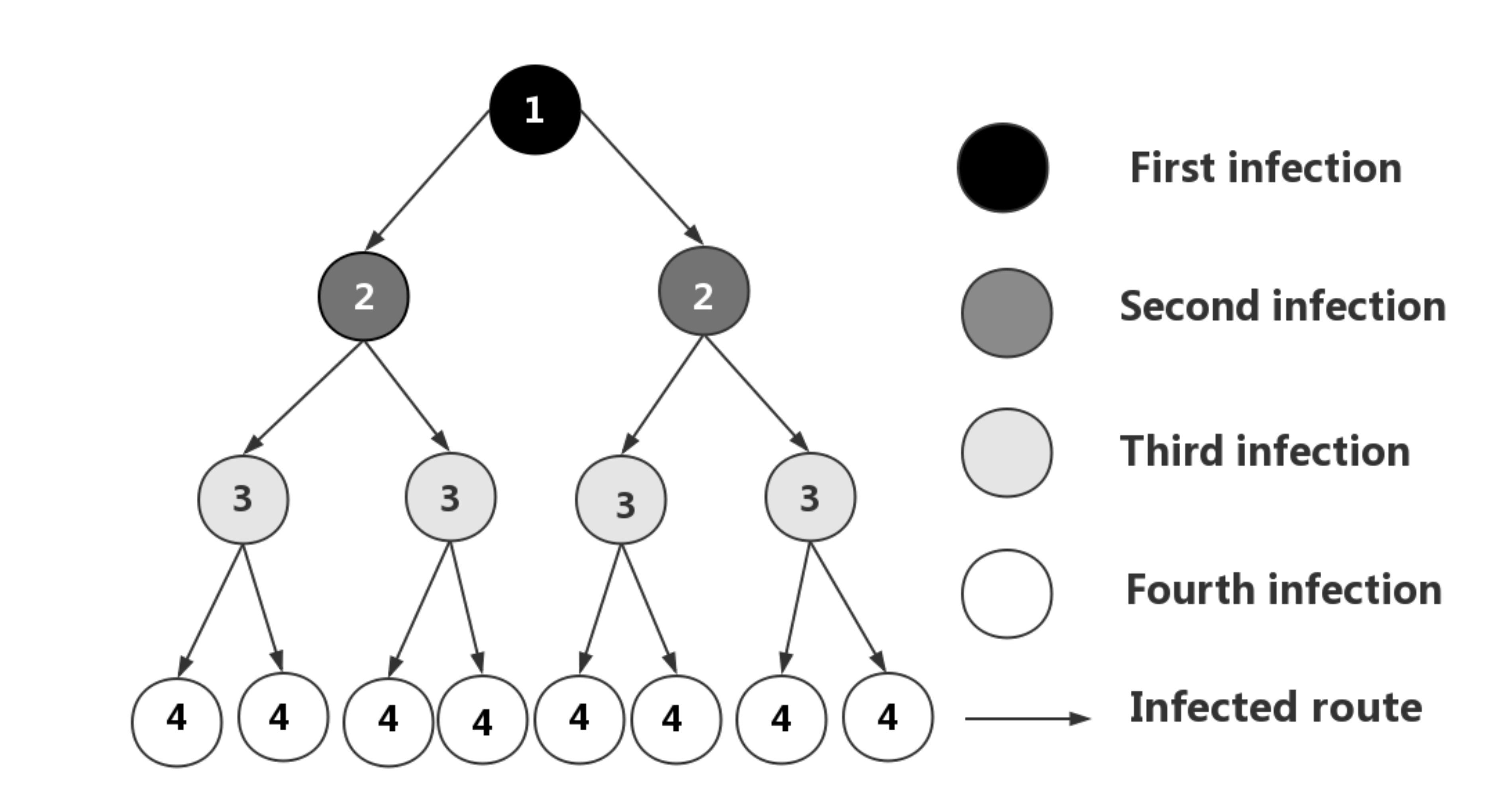}
  \caption{Tree based broadcast network}
\end{figure}

~\\In tree based broadcast network, message could be initialed from any tree branch node. A binary tree node will pass the message to its parent and two children from the original node. Any neighborhood receives the message, the message will be forward to other neighborhoods except the incoming one. For example, original node's parent receives the broadcast message, it will forward the message to its parent node and another leaf node. If the original node's child receives the message, the message will be forward to both its children leaves.

~\\However, tree based network also has its disadvantage compare to gossip protocol\cite{ref_article1}: any single node failure will block the message propagation, because there is only one way for message to travel. To fix this issue, redundancy is required to add in this tree based broadcast network design.

\section*{Tree based network with cluster redundancy}

\begin{figure}[htbp]
  \includegraphics[width=\textwidth]{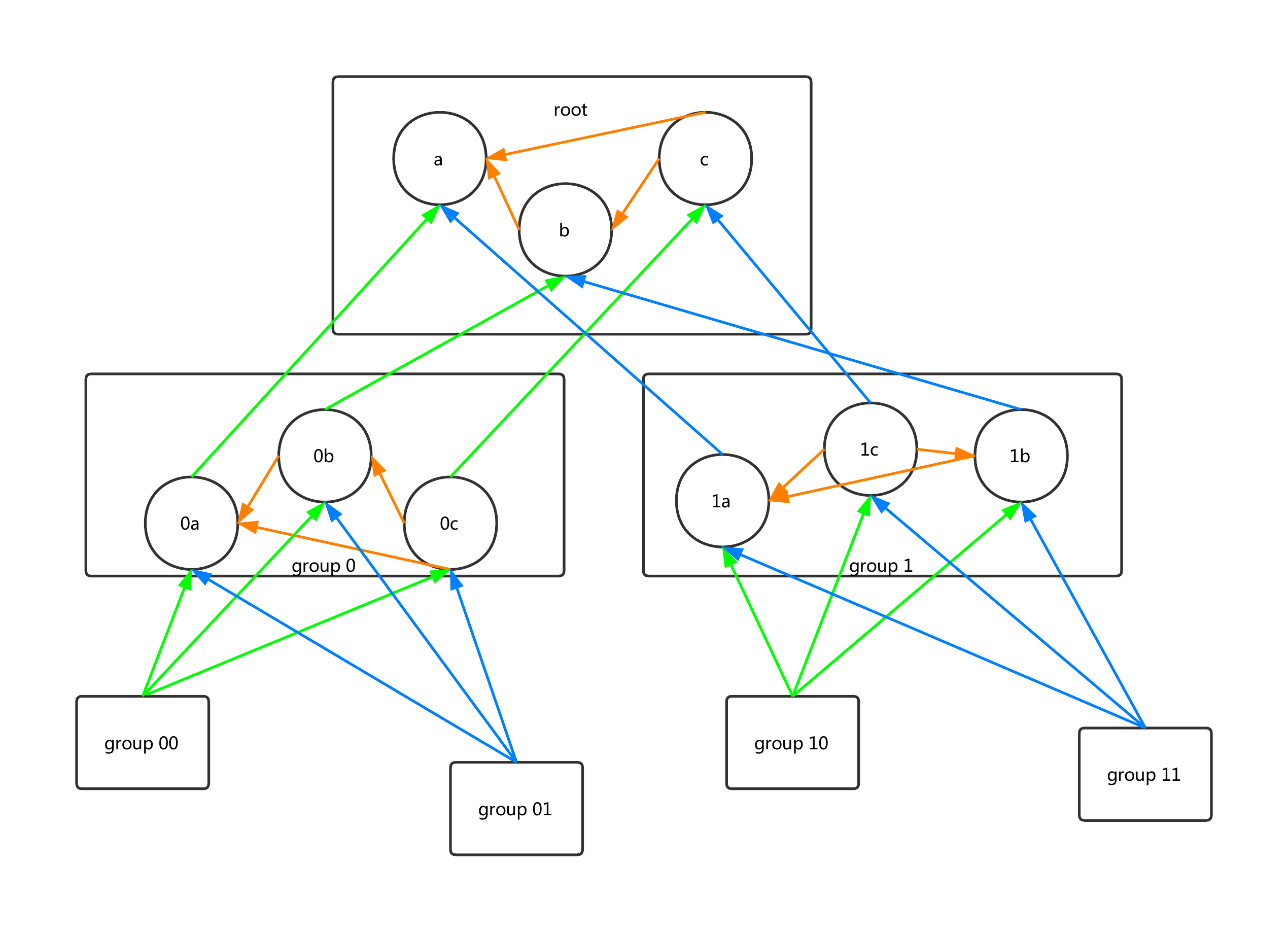}
  \caption{Tree based broadcast network with cluster}
\end{figure}

~\\The basic idea of adding redundancy to the tree based network, is to extend the single node to a cluster group. In Figure 3 for the experiment purpose we assume a group contains 3 nodes. Inside the group, the nodes connect to each other. The nodes within one group can be located in different data center world wide. The broadcast message is passed to the group buddies if any node receives message from either parent or child.

~\\Broadcast storm will happen as each message is hand to not only parent and children but also the group buddies. To prevent broadcast storm the node is required to remember the received message id. When the same message incomes, the node will refuse to forward it again.

~\\The benefit of this design is it allows multi paths of routing from parent to child. If the direct connection between the parent and child is very slow or interrupted, the connection to buddy could be relatively fast. This relay could speed up the message propagation and ensure the reliable of whole tree network.

\section*{Metric}

For the gossip network, there are three key parameters: the network size, the number of contacts, and the message hotness decaying rate. Those parameters will impact the spreading speed. The tree network will reach the 100\% convergence within fixed cycles, which is more efficiency.

~\\In ideal situation, the gossip protocol and the tree based routing provides the similar propagation rate. The total nodes affected within N cycles in gossip protocol is $2^{N - 1}$ and in binary tree based routing is $2^{N}-1$, which is quite similar growth rate.

~\\However, in reality, gossip network can not archive this rate because of nodes' duplicated contact. Let's say each node has enough contacts and the message hotness decaying rate is zero. As the message spreading showing in Figure 4, node may talk to the contact which has already heard the message before. It slows down the whole network propagation, no matter the network size.

\begin{figure}[htbp]
  \begin{center}
  \includegraphics[width=1\textwidth]{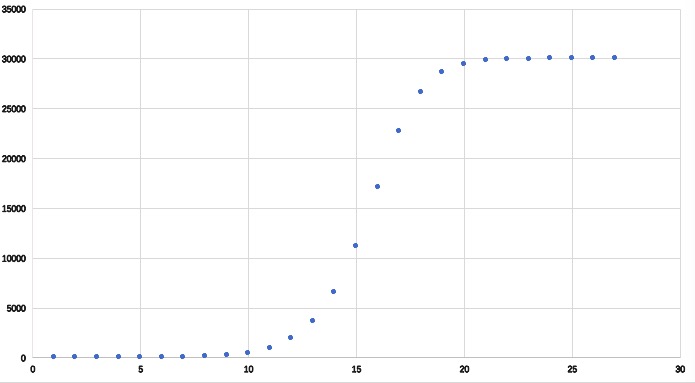}
  \caption{Gossip protocol simulation}
  \end{center}
\end{figure}
~\\In tree based routing the nodes are organized in structure, it's a trade off between pure distributed system and centralized system. The node message always sends to unheard nodes in the next cycle. The convergence to all the nodes will be faster comparing to gossip protocol.

\section*{Evaluation}

Tree based network routing plays quite similar way as gossip protocol did in blockchain broadcasting. Gossip protocol is simple, rousted, reliable, but sometimes it takes longer time to reach the 100\% convergence. Tree based network is easy to understand, less traffic wasted, the disadvantage is obvious: the single node failure issue. Tree cluster network can fix this, although it makes the implementation a bit more complex, however it combines the advantages of both gossip protocol and tree network.This makes the following points a reality. Firstly,  improving the integrating capability of P2P network. Secondly, this network Improves the efficiency of message transmission and avoids power consumption.Thirdly, Reducing the network bandwidth occupancy and improving the broadband speed.

\section*{Related work}

While we were working on the blockchain performance paper: Improve Blockchain Performance using Graph Data Structure and Parallel Mining\cite{ref_article3}, it's found that the blockchain performance has its bottle neck not only in the chain data structure, but also in the P2P network. Comparing to the blockchain, P2P technology is a big issue as the diversity of all kinds network situations. Lots of problem in reality would affect the P2P network performance, generally blockchain performance is limited. Oppositely, people believed improving consensus algorithm can increase blockchain performance, which is not true as expected. To improve blockchain's overall performance, P2P network is the critical part to work on.

~\\In Seele's second yellow paper: An Accelerated Method for Message Propagation in Blockchain Networks\cite{ref_article4}, the performance requirement for a blockchain P2P network is affirmed. To build a high performance, a fast P2P network is necessary component.

\section*{Acknowledgment}

The authors thank Brahma OS and the team advisors' help on information collection and idea pitching.

~\\This work was supported in part by the National Natural Science Foundation of China under Grant No. 61701418, in part by Innovation Projects of The NextGeneration Internet Technology under Grant NGII20170301。

\end{document}